\documentclass[final]{svjour2}

\smartqed

\usepackage{graphicx}
\usepackage{rotating}
\usepackage{enumitem}
\usepackage{amssymb}
\usepackage{mathptmx}
\usepackage[numbers]{natbib}

\makeatletter
\journalname{Journal of Low Temperature Physics}

\bibpunct{}{}{,}{s}{}{,}

\begin{document}

\newcommand{\hdblarrow}{H\makebox[0.9ex][l]{$\downdownarrows$}-}
\title{Study of Cosmic Ray Impact on \textit{Planck}/HFI Low Temperature Detectors}

\author{A. Miniussi$^1$ \and J.-L. Puget$^1$ \and W. Holmes$^2$ \and G. Patanchon$^3$ \and A. Catalano$^4$ \and Y. Giraud-Heraud$^3$ \and F. Pajot$^1$ \and M. Piat$^3$ \and L. Vibert$^1$}

\institute{ $[1]$ \textbf{IAS}, Bt. 121, Universite Paris Sud-11, F-91405 Orsay, France \\ Tel.: +33 (0)169858696  \\ \email{antoine.miniussi@ias.u-psud.fr} \\
$[2]$ \textbf{JPL}, California Institute of Technology, Pasadena, CA, USA, $[3]$ \textbf{APC} 10, rue A. Domon et L. Duquet 75205 Paris Cedex 13, $[4]$ \textbf{LPSC} 53, rue des Martyrs 38026 Grenoble Cedex, France}

\date{09.13.2013}

\maketitle

\begin{abstract}

After the focal plane of the HFI instrument of the Planck mission (launched in May 2009) reached its operational temperature, we observed thermal signatures of interactions of cosmic rays with the Planck satellite, located at the L2 Lagrange point. When a particle hits a component of the bolometers (e.g. thermometer, grid or wafer) mounted on the focal plane of HFI, a thermal spike (called glitch) due to energy deposition is measured. Processing these data revealed another effect due to particle showers of high energy cosmic rays: High Coincidence Events (HCE), composed of glitches occurring coincidentally in many detectors followed by a temperature increase from the nK to the $\mu$K. A flux of about 100 HCE per hour has been estimated.
Two types of HCE have been detected: fast and slow.  For the first type, the untouched bolometers reached, within a few seconds, the same temperature as those which were "touched". This can be explained by the storage of the energy deposited in the stainless steel focal plane. The second type of HCE is not fully understood yet. These effects might be explained by an extra conduction due to the helium released by cryogenic surfaces and creating a temporary thermal link between the different stages of the HFI.


\end{abstract}

\section{Introduction}

The \textit{Planck} satellite, launched in May 2009 from Kourou (French Guyana), reached the second Sun-Earth Lagrangian point (L2) in July 2009. Since this date, and until January 2012, the \textit{Planck} mission\cite{planck_coll_1} has mapped the entire sky five times. The mission is composed of two instruments: the High Frequency Instrument (HFI) and the Low Frequency Instrument (LFI). The HFI is composed of 54 Neutron Transmutation Doped (NTD) germanium bolometers\cite{ini_res_boloplanck} in a focal plane cooled to a temperature of 100 mK \cite{cryo_syst}. This 0.1K plate also carries two germanium thermometers and two active thermal controls (PID).

Even though the HFI is mapping the sky with an unprecedented photon Noise-Equivalent Power (NEP) of $\sim$ 10$^{-17}$ W/Hz$^{1/2}$, HFI bolometers and thermometers record a large number of thermal spikes due to energy deposit from cosmic rays, called \textit{glitches}\cite{hfi_patanchon}.

Cosmic rays are emitted by different sources (stars, supernovae, etc.) and are mainly composed of protons (86\%) and helium nuclei (11\%)\cite{cr_flux}. For particles of energy lower than 10 GeV, the flux is modulated by solar wind. Above this limit, the spectrum can be described by a power law distribution in energy, E$^{-2.5}$ \cite{cr_lowenergy}. At Lagrangian point L2, Planck is mostly affected by protons interacting with the different materials composing the satellite. The contribution of solar particles is negligible outside of major solar flares.

These particles can interact in two different ways with matter:
\begin{itemize}[noitemsep,nolistsep]
\item energy deposit: at relevant energies of particle, the deposited energy primarily depends on the material thickness and atomic number. The deposited energy is dominated by electronic stopping power.
\item production of secondary particles: the particle interacting with matter produces a secondary shower of lower energy particles.
The multiple particles produced have a lower energy, and therefore deposit energy more efficiently when the primary particle has interacted somewhere in the spacecraft with material.
\end{itemize}

\begin{figure}
	\begin{center}
		\includegraphics[width=1\linewidth,keepaspectratio]{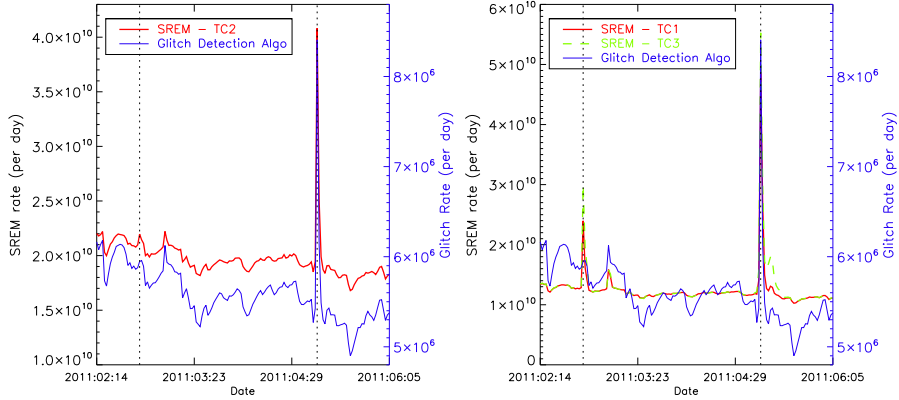}
	\end{center}
	\vspace{-1.2em}
	\caption{Comparison of TC1, TC2 and TC3 diodes and bolometer glitch rates for the same period of time (about 150 days) which includes two solar flares, designated in by dashed lines. \textbf{Left} : the TC2 diode and glitch rates are very well correlated. The solar flare in June 2011 clearly appears while the one in March 2011 is negligible. \textbf{Right} :  TC1 and TC3 diodes both record the solar flares meaning that the first one is composed of fewer energetic particles.}
	\vspace{-1.2em}
	\label{srem_glitch}
\end{figure}

The ESA Standard Radiation Environment Monitor (SREM) \cite{srem}, located in the Planck spacecraft measures the low energy particle flux at L2 which is mainly composed of protons (electrons are only typically 1 \% of the cosmic rays). The SREM is composed of 3 diode sensors: TC1, TC2 and TC3. TC2 is shielded with 1.7 mm of Aluminum and 0.7 mm of Tantalum which makes it sensitive to ions and protons with energies greater than 39 MeV. TC1 and TC3 are respectively shielded with 1.7 mm and 0.7 mm of Aluminum.
The measurement of the cosmic ray flux observed on bolometers is very well correlated when compared to the TC2 diode as shown in fig.~\ref{srem_glitch}. TC1 and TC3 diodes, less shielded, are able to record solar flares not seen in TC2 and the bolometer. This comparison demonstrates that only protons of an energy above $\sim$ 40 MeV can reach the focal plane and interact with bolometers. Lower energy particles are stopped by the material surrounding the HFI bolometers.

\section{The High Coincidence Event (HCE)} \label{chap:hce}

\begin{figure}
	\begin{center}
		\includegraphics[width=0.6\linewidth,keepaspectratio]{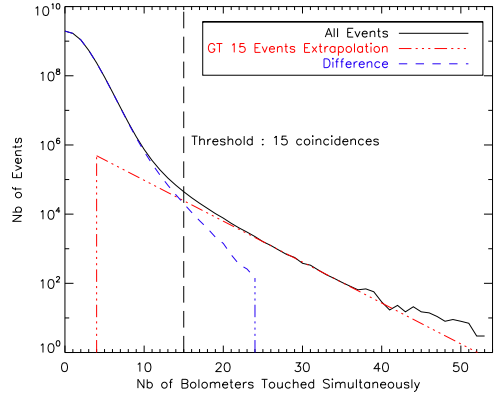}
	\end{center}
	\vspace{-1.2em}
	\caption{Coincidence histogram (solid line) processed for the entire mission ($\sim$ 850 days), based on the results of the glitch detection algorithm for a 15 ms bin (or 3 samples) \cite{hfi_patanchon}. The red dashed line is the extrapolation based on histogram values greater than the 15 coincidences threshold. The blue dashed line is the difference between all events and the extrapolation.}
	\vspace{-1.2em}
	\label{histoglitch}
\end{figure}

The histogram in Fig.~\ref{histoglitch} is built by counting the number of glitches occurring coincidentally in different detectors (called "touched" bolometers) of the 0.1 K plate for a 15 ms bin. The histogram is clearly divided in two parts. Below 12 touched bolometers, the histogram is dominated by random coincidences. Above this limit, other types of events, not due to fortuitous cosmic rays, prevails and are categorized as High Coincidence Events or HCE. We set the detection threshold at 15 in order to only select real coincidences linked with HCE.

The integral of the extrapolation based on the histogram values for events greater than 15 is about 100 HCE per hour but only 2.10$^{-2}$ \% (one every two days) are above 3$\sigma$ bolometer noise.

The rate of these HCE, computed for the entire mission, corresponds to the expected rate of cosmic rays of several TeV interacting with the spacecraft \cite{cr_flux}.

Following the detection of the coincident glitches, we observe a temperature increase of the bolometer plate up to tens of $\mu$K seen in bolometer signals, as well as in the thermometers. Thus, the HCE don't only affect bolometers but rather the entire 0.1 K plate.
This effect is also observable on the thermal regulator which reduces its power in order to counterbalance the 0.1 K plate heating up. Unfortunately, HCE events are too fast compared to the PID time constant which is only able to reduce by 10\% the maximum amplitude of the largest events (greater than 1.5 uK).

For events greater than the bolometer noise ($\sim$ 0.8$\mu$K) one can measure the thermal amplitudes and the time constants. Thus, two types of HCE are observed throughout the mission: "slow" and "fast". They are mainly separated by their rise time constant and the number of touched bolometers. Only the decay time constant of $\sim$ 30 min is similar and clearly linked with the thermal time constant between the 0.1 K plate and the dilution plate \cite{piat}.

\subsection{Fast HCE}
The fast HCE are described by these characteristics : 
\begin{itemize}[noitemsep,nolistsep]
\item rise time constant : $\sim$ 5 s 
\item coincident glitches are geometrically grouped and occur in several, but not all bolometers
\end{itemize}

\begin{figure}
\centering
\begin{minipage}{0.5\textwidth}
  \centering
  \includegraphics[width=1\linewidth,keepaspectratio]{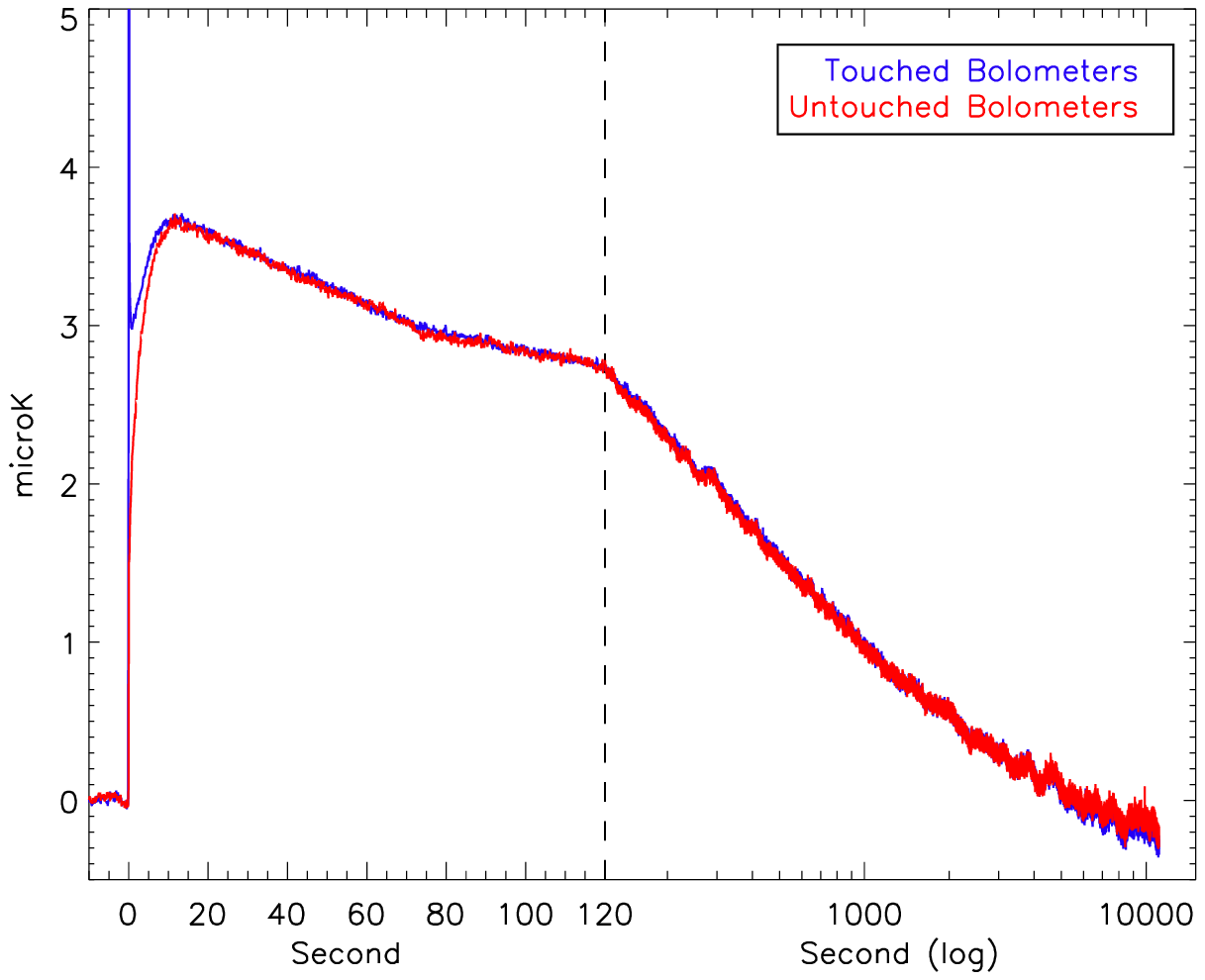}
  \label{stack_court}
\end{minipage}%
\begin{minipage}{0.5\textwidth}
  \centering
  \includegraphics[width=1\linewidth,keepaspectratio]{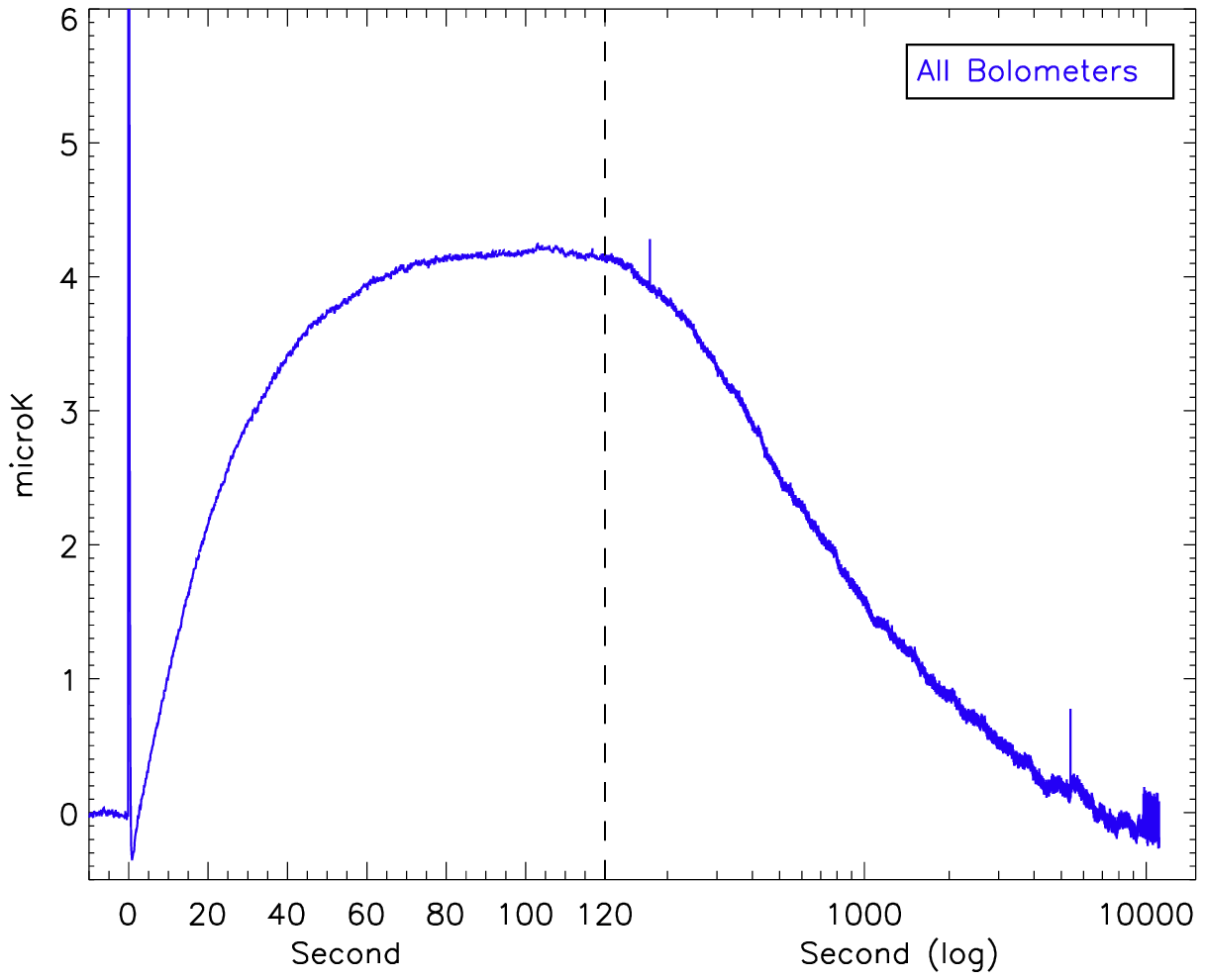}
  \label{stack_long}
\end{minipage}
\vspace{-1.2em}
\caption{\textbf{Left} : Stacking of fast HCE based on events greater than 1 $\mu$K. Touched and untouched bolometers are dissociated. \textbf{Right} : Stacking of all bolometers for slow HCE greater than 1 $\mu$K.}
\vspace{-1.2em}
\label{stack_bolo}
\end{figure}

As shown in Fig.~\ref{stack_bolo} the temperature of untouched bolometers follow the temperature rise of those which were touched. Thus, the thermal effect of HCE is linked with the entire 0.1K plate, and not only with the bolometers.

The interaction process can be formulated as follows: a high energy particle interacts with the HFI/LFI focal plane unit and creates a particle shower. Secondary low-energy particles touch the bolometers and deposit energy in the 0.1 K plates which heats it up. The short distance between the HFI/LFI box and the 0.1 K plate allows only a narrow particle shower. This explains that only a subset of neighboring bolometers are touched, but all bolometers see the associated heating of the focal plane over longer time scales.

The heat capacity and the diffusivity of the stainless steel 0.1 K plate allows the energy of the particles to be stored in it and released with a time constant of a few seconds which is coherent with the rapid rise typical of fast HCE. To heat the 0.1 K plate by $\sim$ 1 $\mu$K, a deposited energy of $\sim$ 1 TeV is necessary.

\subsection{Slow HCE}
The slow HCE are characterized as follow : 
\begin{itemize}[noitemsep,nolistsep]
\item rise time constant : $\sim$ 30 s
\item nearly all or all bolometers are touched
\item a small temperature ($\sim$ 0.5 $\mu$K) decrease appears after the coincident glitch and before the temperature rise (Fig.~\ref{stack_bolo})
\item precursor glitches have different time constants from all other types observed so far
\end{itemize}

For this type of HCE, the particle shower size is larger as all bolometers are touched. In comparison with fast HCE, the primary particle must have interacted in the spacecraft further from the 0.1 K plate.

The slow thermal rise of the bolometer plate temperature, lasting about one minute, implies a thermal transfer from warmer elements as the energy cannot be stored in the 0.1 K plate without being detected. This could be explained by the creation of a temporary thermal link between the bolometer plate and the 1.6 K environment.
Furthermore, the temperature time stream goes through a short cool down period for about 2 seconds after the shower is detected. This can only come from a temporary thermal efficient link between the bolometer plate and the colder dilution plate. We thus have evidence for two temporary thermal links.
The only hypothesis found so far to explain these temporary thermal links is molecular conduction of helium atoms, briefly freed by the particle shower, from surrounding surfaces. Helium atoms could come from the helium tanks, or very small leaks from the cryostat.

\begin{figure}
\begin{center}
\includegraphics[width=0.6\linewidth,keepaspectratio]{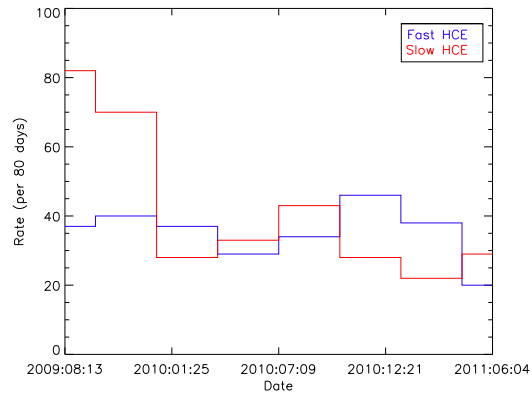}
\end{center}
\vspace{-1.2em}
\caption{Flux of slow and fast HCE greater than 1 $\mu$K for 80 day bins. For each HCE type, we detected 171 events greater than 1 $\mu$K}
\vspace{-1.2em}
\label{taux_ele}
\end{figure}

As shown in Fig.~\ref{taux_ele}, the rate of slow HCE has decreased during the Planck mission, while the number of fast HCE has remained constant. As HCE are caused by high energy cosmic rays (see \ref{chap:hce}), exceeding the effects of solar modulation, none of the HCE are linked with solar activity.
The phenomenon can be accounted for by a decrease in the amount of helium trapped on the 0.1 K system surfaces through the degassing holes.

\section{Conclusions}

HCE will be investigated further with an experiment at the Institut d'Astrophysique Spatiale (IAS), in collaboration with JPL, to simulate the effect of helium thermal exchanges between plates at different temperatures. Moreover, data processing is still ongoing and should provide us more information about HCE. 

The High Frequency Instrument is a technological success. The information gathered on the HCE is a unique source of information on the behavior of 100 mK systems in space.
Future space missions will aim at NEP per-detector of $\sim$ 10$^{-18}$ W/Hz$^{1/2}$ or less which will lead to a greater impact of HCE and glitches on data. Thus, future missions will have to be designed taking into account such thermal effects as cosmic ray, vibration, etc.

Before flight, cosmic ray impact on detectors and conductance and capacitance measurements of all sensitive elements should be thermally studied (detectors and any thermally coupled material). The helium, present in most cryogenic systems, might also play a role in thermal conduction. Behavior and localization of helium should be studied to avoid any important thermal exchange. 
Finally, in order to reduce the impact of HCE-like thermal events, the thermal regulation could be more reactive with a shorter time constant, but with thermometers less sensitive to particles (a difficult trade-off).

In order to validate the next generation of sub-kelvin space detectors, tests with energetic space and sea-level particles have to be run. They have been completed with a Transition Edge Sensor (TES) bolometer using an alpha particle source ($^{241}$Am) at APC (see Joseph Martino's paper in this journal). Others are in progress with the Kinetic Inductance Detector (KID) at the Neel Institute \cite{kid}.

\end{document}